\documentclass[11 pt]{article}
\usepackage{amsfonts,latexsym, amssymb, amsmath} 
\usepackage{graphicx,subfigure,epsfig} 

\newcounter{cst}

\newcounter{cexp}

\renewenvironment{abstract}{\begin{center} {\bf Abstract} \\
\end{center} }{\smallskip}

\begin{document}

\def\dsp{\displaystyle}

\newcommand\mc{\spadesuit}
\newcommand\gk{\clubsuit}

\def\be{\begin{equation}}
\def\ee{\end{equation}}

\def\bea{\begin{eqnarray}} 

\def\eea{\end{eqnarray}}

\def\beqsys {\be\ba \left \{ \begin{array}{l}}
\def\eeqsys {\end{array} \right . \ea\ee }

\def\beqsysno {\be\ba \left \{ \begin{array}{l}}
\def\eeqsysno {\end{array} \right . \ea\ee}


\def\app{{\DD}}
\def\appm{{\DD_m}}

\def\BB{{\cal B}}
\def\Bapp{\BB_\DD}

\def\C{\mathbb{C}}
\def\CC{{\cal C}}
\def\cun{{{\cal C}^1}}
\def\czero{{{\cal C}^0}}
\def\dklnp{\delta_{K,L}^{n+1}}
\def\Dhcarreuklnp{{{\dklnp (h^2(U))}}}
\def\DD{{\cal D}}
\def\demi{\frac{1}{2}}
\def\Dfcarreuklnp{{{\dklnp (f^2(U))}}}
\def\Dfuklnp{{{\dklnp (f_\mu(U))}}}
\def\Dguklnp{{\dklnp (g(U))}}
\def\Dhuklnp{{{\dklnp (h(U))}}}
\def\diam{\hbox{\rm diam}}
\def\div{{\rm div}}
\def\Dkaklnp{{{\dklnp (\mu k_a(U))}}}
\def\Dkwklnp{{{\dklnp (k_w(U))}}}
\def\dkl{d_{\kl}}
\def\dO{\partial \O}
\def\Dpcuklnp{{\dklnp (p_c(U))}}
\def\DPdeklnp{{\dklnp (Q)}}
\def\dt{{\delta t}}
\def\dtn{{\dt}^n}
\def\Duklnp{{\dklnp (U)}}

\def\EEdnp{{\cal E}_d^{n+1}}
\def\EEsnp{{\cal E}_s^{n+1}}
\def\eps{\varepsilon}
\def\EEunnp{{\cal E}_1^{n+1}}
\def\EEdenp{{\cal E}_2^{n+1}}
\def\epskl{{\varepsilon_{\kl}^{n+1}}}

\def\fapp{f_{\matt,\dt}}
\def\FF{{\cal F}}
\def\fknp{f_K^{n+1}}

\def\GG{{\cal G}}
\def\GGdeklnp{{\Phi_\mu}_{2,K,L}^{n+1}}
\def\GGklnp{{\Phi_\mu}_{K,L}^{n+1}}
\def\GGunklnp{{\Phi_\mu}_{1,K,L}^{n+1}}
\def\grad{\nabla}

\def\half{{\frac 1 2}}
\def\hun{H^1(\O)}

\def\ie{{\it i.e. }}
\def\intO{\dsp{\int_{\O}}}
\def\intQ{\dsp{\int_0^T\!\!\!\int_{\O}}}
\def\intT{\dsp{\int_0^T}}
\def\intxi{\dsp{\int_{\O_\xi}}}

\def\kaklnp{k_a(U_{\kl}^{n+1})}
\def\ki{\,\mathbf{1} \!\!\!\!\;{1}}
\def\kl{{K|L}}
\def\kwklnp{k_w(U_{\kl}^{n+1})}

\def\lap{\Delta}
\def\lde{{L^2(\O \times (0,T))}}
\def\ldehun{{L^2(0,T,H^1(\O))}}
\def\linfty{L^{\infty}(\O \times (0,T))}

\def\mK{m_K}
\def\mmin{{M_0}}

\def\n{\mathbf{n}}
\def\N{\mathbb{N}}
\def\E{\mathbb{E}}
\def\bB{{\bf  B }}
\def\by{{\bf y} }
\def\bS{{\bf S}\, }
\def\bz{{\bf z} }
\def\bZ{{\bf Z} }
\def\bL{{\bf L} }
\def\bM{{\bf M} }
\def\bl{{\bf l} }
\def\bmm{{\bf m} }
\def\bW{{\bf W}}
\def\sigb{\bar{\sigma}}
\def\taub{\bar{\tau}}
\def\LL{ \mathcal{L}}

\def\EE{{\cal E}}
\def\TT{{\cal T}}
\def\UU{{\cal U}}

\def\pardt{{\partial_t}}
\def\pardB{{\partial_B}}
\def\pardy{{\partial_y}}
\def\pards{{\partial_S}}
\def\pardz{{\partial_z}}
\def\pardY{{\partial_Y}}
\def\pardss{{\partial_{SS}}}
\def\pardsss{{\partial_{S}^3}}
\def\pardssss{{\partial_{S}^4}}
\def\pardzz{{\partial_{zz}}}
\def\pardsz{{\partial_{Sz}}}
\def\pardyy{{\partial_{yy}}}
\def\pardYY{{\partial_{YY}}}

\def\ep{{\varepsilon}}
\def\epin{\frac{1}{\ep}}
\def\epins{\frac{1}{\sqrt{\ep}}}

\def\nkl{\n_{K,L}}

\def\O{\Omega}
\def\on{[\! [ 0,N ]\! ]}
\def\onp{[\! [ 0,N+1 ]\! ]}
\def\OO{{\cal O}}

\def\Papp{P_{\DD}}
\def\Pappm{P_{\DD_m}}
\def\Pc{{p_c}}
\def\Pdekn{Q_K^n}
\def\Pdeknp{Q_K^{n+1}}
\def\Pdeln{Q_L^n}
\def\Pdelnp{Q_L^{n+1}}
\def\Pg{{p_g}}
\def\Qg{{q_g}}
\def\phi{\varphi}
\def\Pkn{P_K^n}
\def\Pknp{P_K^{n+1}}
\def\Pln{P_L^n}
\def\Plnp{P_L^{n+1}}
\def\problem{(\ref{eq1})-(\ref{moyennep}) }
\def\ps{\cdot}
\def\psiapp{\psi_{\DD}}
\def\psiappm{\psi_{\DD_m}}
\def\psikn{\Psi_K^{n}}
\def\psiknp{\Psi_K^{n+1}}
\def\psilnp{\Psi_L^{n+1}}

\def\q{\mathbf{q}}
\def\Q{\mathbb{Q}}
\def\quart{\frac{1}{4}}

\def\R{\mathbb{R}}

\def\scheme {(\ref{initcond})-(\ref{moyennenulle}) }
\def\size{\hbox{\rm size}}
\def\smoins{\,{\underline{s}}}
\def\smoinsknp{\smoins_K^{n+1}}
\def\splus{\,{\overline{s}}}
\def\splusknp{\splus_K^{n+1}}
\def\sumK{\dsp{\sum_{K\in\matt}}\,}
\def\sumL{\dsp{\sum_{L\in \NN_K}}}
\def\sumN{\dsp{\sum_{n=0}^{N}}\,}
\def\tnp{t^{n+1}}
\def\tendto{\to}
\def\tkl{\tau_{\kl}}
\def\tn{t^n}
\def\troisquart{\frac{3}{4}}

\def\uapp{{u_{\DD}}}
\def\uappm{{u_{\DD_m}}}
\def\udiscret{(\uknp)_{K \in\matt, n \in \on}}
\def\ukn{U_K^n}
\def\uknp{U_K^{n+1}}
\def\ulnp{U_L^{n+1}}
\def\ukzero{{U_K^0}}
\def\unnp{[\! [ 1,N+1 ]\! ]}
\def\uklnp{U_{\kl}^{n+1}}
\def\ubklnp{{\bar U_{\kl}^{n+1}}}

\def\weakto{\wtendto}
\def\wtendto{\rightharpoonup}

\def\Z{\mathbb{Z}}

\def\cknp{c_K^{n+1}}

\catcode`\@=11
\newif\ifproofmode
\proofmodetrue
\def\labelcour{nolabel}

\def\theequation{\thesection.\arabic{equation}}

\def\@endtheorem{\hfill\ifproofmode\rlap{\tiny \kern5mm
\labelcour}\fi\endtrivlist}

\catcode`\@=12 \proofmodefalse
\title{European Option Pricing with Transaction Costs and Stochastic
Volatility: an Asymptotic Analysis}

\author{R. E. Caflisch\footnote{Department of Mathematics, UCLA, USA, rcaflisch@ipam.ucla.edu}$\;$
G. Gambino\footnote{Department of Mathematics, University of Palermo, Italy, gaetana@math.unipa.it}$\;$
M. Sammartino\footnote{Department of Mathematics, University of Palermo, Italy, marco@math.unipa.it} $\;$
C. Sgarra\footnote{Department of Mathematics, Politecnico di Milano, Italy, carlo.sgarra@polimi.it}
}


\maketitle

\medskip

\begin{abstract}
In this paper the valuation problem of a European call option in presence
of both stochastic volatility and transaction costs is considered.
In the limit of small transaction costs and fast mean reversion,
an asymptotic expression for the option price is obtained.
While the dominant term in the expansion it is shown to be the classical
Black and Scholes solution, the correction terms appear at $O(\varepsilon^{1/2})$ and
$O(\varepsilon)$. The optimal hedging strategy is then explicitly obtained for
the Scott's model.
\end{abstract}

\section{Introduction}
\setcounter{equation}{0}

The intrinsic limitations of the Black-Scholes model in describing real
markets behavior are very well known. Among the main assumptions underlying
that model the assumptions of constant volatility and no transaction costs are
probably the most relevant. In this paper we are going to consider the pricing
problem of a European option in a model in which both proportional transaction
costs are taken into account and the volatility is assumed to evolve according
to a stochastic process of the Ornstein--Uhlenbeck type.
To analyze this situation we shall follows a utility maximization
procedure, following the seminal paper of M.H.A.Davis, V.G.Panas and
T.Zariphopoulou \cite{DPZ}.

If one uses the following utility function ${\cal U}$:
$$
{\cal U}(x)= 1-\exp{(-\gamma x)} \; ,
$$
where $\gamma$ expresses the risk aversion of the investor,
one gets, as the result of this analysis, a non linear PDE for the
expected value of the utility--maximized wealth held in the
underlying asset of the option.

At this point we shall make two assumptions.
First, that the process
driving the volatility is fast mean--reverting with speed
$O(\ep^{-1})$.
Second, that the transaction costs are {\em very} small,
$O(\ep^{-2})$.

The pricing of a European option in presence of small transaction
costs was considered in \cite{WW}, where a correction term to the
Black\&Scholes pricing formula was derived. This correction term
was found to be order $2/3$ in the pricing cost. Moreover in
\cite{WW} was found that the optimal hedging strategy consisted in
not transacting when the process driving the stock price is in a
strip around the classical Black\&Scholes {\em delta}--hedging
formula and in rebalancing the portfolio (selling or buying
stocks) to keep the process inside the strip of no transaction.
The width of the no transaction strip was found to be order $1/3$
in the transaction costs. In \cite{WW} the volatility was supposed
to be constant. More Recently V.I.Zakamouline studied optimal hedging
of European options with transaction costs via a utility optimization
approach in (\cite{Zaka2006}).

The pricing of  a European option with fast mean--reverting
stochastic volatility was considered in in a series
of papers (see e.g. \cite{FPS}, \cite{FPSS03} and \cite{FPSS99}).
In the above mentioned papers the authors found the pricing formula
whose leading order term is the classical Black\&Scholes formula
with averaged volatility. The correction term was order the square root
of the characteristic time scale of the process driving the volatility.
An optimal consumption-investment problem has been investigated in a
paper by M.Bardi, A.Cesaroni and L.Manca (\cite{BCM}) where a rigorous
asymptotic analysis is performed and where the solution is characterized
in the limit of fast volatility dynamics.

More recently, M.C.Mariani, I.SenGupta and P.Bezdek (\cite{MSB}) proposed
a numerical approximation scheme for European option prices in
stochastic volatility models including transaction costs  based on
a finite-difference method. The stochastic volatility dynamics
considered there is a slight generalization of that proposed by
Hull and White \cite{HW1987}, since they consider a drift coefficient
which is a general (regular) deterministic function of both the
time and the underlying asset price, while their diffusion coefficient
is linear in the instantaneous volatility.

In the present paper we propose a different approximation method
for European option pricing in stochastic volatility models with
transaction costs, based on an asymptotic analysis which follows the
approach pioneered by J.-P.Fouque, G.Sircar and R.Papanicolaou (\cite{FPS}).
The model we consider for the stochastic volatility dynamics is of
Ornstein--Uhlenbeck type with a constant diffusion coefficient.
This model has been originally proposed by Stein and Stein
\cite{SS1991}. We provide closed-formulas for European option prices
in the limit of fast volatility and small transaction costs.

The plan of the paper is the following: in Section 2 we introduce
the multidimensional stochastic process for the dynamic of a
portfolio of the writer of a European option. By solving a stochastic
control problem, we obtain the Hamilton-Jacobi-Bellman equation.
In Section 3, the asymptotic analysis is done assuming small
transaction costs and fast mean reverting volatility. In section 4
the price of the option is computed and in section 5 the numerical
results are provided and some conclusions outlined.

For the reader's convenience, in Appendix A the source term of the equation
obtained through the asymptotic analysis at $O(\varepsilon)$ is
calculated; in Appendix B the averages with respect to the
Ornstein-Uhlenbeck invariant measure using the model introduced by
L.Scott (\cite{CS}) for the stochastic volatility are given; finally, in Appendix
C the derivatives, which appears in the obtained corrected pricing
formula, with respect to the stock price of the classical Black
and Scholes model are recalled and collected.

\section{Option pricing via utility maximization}
\setcounter{equation}{0}

We suppose to have the following multidimensional stochastic
process:

\bea
d\bB &=& r\bB dt -(1+\lambda)\bS d\bL + (1-\mu)\bS d\bM \\
d\by &=& d\bL-d\bM \\
d\bS &=& \bS \left( \alpha dt + f(\bz) d\bW \right)\\
d\bz &=&  \xi(m-\bz)dt+\beta\left(\rho d\bW+
\sqrt{1-\rho^2} d\bZ \right)\; .
\end{eqnarray}

\noindent In the above equations $\bB$ and $\bS$ are the risk-free
(the "Bond") and the risky asset (the "Stock") respectively,
$r$ is the risk--free interest rate,
$\alpha$ is the drift rate of the stock,
$\lambda$ and $\mu$ are the (proportional) cost of buying and selling
a stock, $f$ is the volatility function, which we shall suppose to
depend on the stochastic variable $z$, which is sometimes called
the volatility driving process. $\bL$ and $\bM$ are the cumulative
number of shares bought or sold, respectively, in the time interval
considered $[0,T]$. We keep the notations introduced in \cite{DPZ}
and \cite{WW}, where the reader can find a detailed justification
for the model for transaction costs just introduced.
In what follows we shall always suppose  $f(z)$ to be a function
bounded away from $0$:

$$
0<m_1\leq f(z)\leq m_2 <\infty,      \qquad \forall z.
$$

\noindent
The process followed by the stochastic variable $z$ is a
Ornstein--Uhlenbeck process with average $m$. The parameter $\xi$
is the rate of mean reversion volatility.

The Brownian motions $\bW$ and $\bZ$ are uncorrelated and $\rho$
is the instantaneous correlation coefficient between asset price and
the volatility shocks. Usually one considers $\rho<0$, i.e. the two
processes are anti--correlated (e.g. when the prices go
down the investors tend to be nervous and the volatility raises).
For more details see \cite{FPS} and \cite{JS}.

We will suppose to deal with trading strategies absolutely continuous
with respect to time, so that:

$$
\bL=\int_0^t \bl ds\,, \qquad  \bM=\int_0^t \bmm ds \; .
$$
Therefore the process we are dealing with can be written in the form:
\bea
d\bB &=& \left[r\bB -(1+\lambda)\bS \bl + (1-\mu)\bS \bmm \right]dt \\
d\by &=& (\bl-\bmm)dt \\
d\bS &=& \bS \left( \alpha dt + f(\bz) d\bW \right)\\
d\bz &=&  \xi(m-\bz)dt+\beta\left(\rho d\bW+
\sqrt{1-\rho^2} d\bZ\right) \; .
\end{eqnarray}

\noindent The final value of a portfolio of the writer of a European option
with strike price $K$, after following the strategy $\pi$ is:
\bea
\Phi_w(T,\bB^\pi(T),\by^\pi(T), \bS(T),\bz(T))&=&
\bB^\pi(T)+I_{(\bS(T)<K)} c(\by^\pi(T),\bS(T))
+ \nonumber \\
&&I_{(\bS(T)>K)}\left[ c(\by^\pi(T)-1,\bS(T))+K\right] \; .
\eea
On the other hand the final value of a portfolio which does not include
the option is simply:
\be
\Phi_1(T,\bB^\pi(T),\by^\pi(T), \bS(T),\bz(T))=\bB^\pi(T)+
c(\by^\pi(T),\bS(T)) \; .
\ee

We can now define the following value functions:
\bea
V_j(B)=\sup_{\pi\in {\TT}}\E \left(\UU \left(
\Phi_j\left(T,\bB^\pi(T),\by^\pi(T),
\bS(T),\bz(T)\right)\right)\right)
\eea
for $j=1,w$. Notice how this value functions depend on the initial
endowment $B$.

Following \cite{DPZ} we now define:
$$
B_j=\inf\left\{B: V_j(B)\geq 0\right\}.
$$

The {\em fair} price of the option $C$ to avoid arbitrage,
i.e. the amount of money that the writer has to receive to accept
the obligation implicit in writing the option,
will therefore be:
\be
C=B_w-B_1 \; . \label{price1}
\ee
For this price the investor would in fact be indifferent between
the two possibilities of going into the market to hedge the option,
or of  going into the market without the option.

We can define the following function that will be useful in the sequel:

\be
\Psi_j(T,\bB^\pi(T),\by^\pi(T), \bS(T),\bz(T)) =
\Phi_j(T,\bB^\pi(T),\by^\pi(T), \bS(T),\bz(T))-\bB^\pi(T).
\ee

\noindent We have to find an equation for $V_j$. In what follows
we shall suppress the index $j$ and denote $V_j$ with $V$.
The problem we are dealing with is a stochastic control problem,
where the control is the trading strategy $\bmm$ and $\bl$.
The equation for $V$ is the Hamilton--Jacobi--Bellman
equation.

\vskip.3cm
\noindent {\bf The Hamilton--Jacobi--Bellman equation:} Suppose we have the following $n$-dimensional controlled stochastic process:

$$
dX= b(t,X,y)dt+\sigma (t,X,y) dW.
$$

\noindent Let us define the performance functional:

$$
J^y(s,x) = \mathbb{E} \left(K(T,X(T))\right).
$$

\noindent In general one can define a performance functional that depends
on the whole trajectory.
In our case we do not need this generality.

The infinitesimal operator associated with the stochastic process
is:
$$
L^y f= \pardt f+b_i\partial_{x_i} f+ a_{ij}\partial_{x_i x_j} f,
$$

\noindent where
$$
a_{ij}=\frac{1}{2}\left(\sigma\sigma^T\right)_{ij}.
$$

If one defines:

$$V=\sup \left\{ J^y : y=y(x) \quad \mbox{is a Markov control}\right\},$$

\noindent then we have the following result according
to the Hamilton-Jacobi-Bellman Theorem: $\sup_y\left\{ L^y V\right\} =0.$

In the present case this turns out to be equivalent to the following otimization problem:

\bea
\max_{0\leq l,m\leq k} \left\{
\left( \pardy V_j -(1+\lambda) S \partial_B V_j\right)l -
\left( \pardy V_j -(1-\mu) S \partial_B V_j\right)m +
\nonumber  \right.\\
\pardt V_j +rB\partial_B V_j +\alpha S \pards V_j +
\xi(m-z) \pardz V_j + \nonumber \\
\left.
\frac{1}{2} [f(z)]^2 S^2 \pardss V_j +\frac{1}{2}\beta^2 \pardzz V_j
+\beta fS\rho\pardsz V_j
\right\} =0.
\eea

\noindent With some analysis, as in \cite{DPZ}, one gets that the above equation
is equivalent to the following equation:

\bea \max \left\{ \left( \pardy V_j -(1+\lambda) S \partial_B
V_j\right),  -
\left( \pardy V_j -(1-\mu) S \partial_B V_j\right),
\nonumber  \right.\\
\pardt V_j +rB\partial_B V_j +\alpha S \pards V_j +
\xi(m-z) \pardz V_j + \nonumber \\
\left. \frac{1}{2} [f(z)]^2 S^2 \pardss V_j +\frac{1}{2}\beta^2
\pardzz V_j +\beta fS\rho\pardsz V_j \right\} =0. \eea

\noindent We now consider the case of the exponential utility function
${\cal U}(x)=1-\exp{(-\gamma x)}$.
We note, just in passing, that this gives for $V_j$ the following
expression:
$$
V_j=1-\inf\left\{\mathbb{E}\left[ \exp{\left(-\gamma B(T)\right)}
\exp{(-\gamma\Psi_j)} \right]\right\},
$$

\noindent where $\Psi_j$ has been previously introduced.

In the above maximization problem let us change the variables passing $V_j\longrightarrow W_j$:
$$
V_j=1-\exp{\left(-\frac{\gamma}{\delta}(B+W_j)\right)}\,,
$$
\noindent where
$$
\delta\equiv\exp{[-r(T-t)]}\; .
$$

\noindent Notice that with the above expression for $V_j$ the price of the
option $C$, as given in (\ref{price1}), now becomes:

\be C=W_1-W_w
\; . \label{price2} \ee

\noindent The maximization problem for $V_j$ is equivalent to the
following minimization problem for $W_j\,$:

\bea \min \left\{ \left( \pardy W_j -(1+\lambda) S \right) \; , \;
\left( -\pardy W_j +(1-\mu) S \right)  \; ,
\right.\nonumber  \\
\pardt W_j -r  W_j +\alpha S \pards W_j +
\xi(m-z) \pardz W_j + \nonumber \\
\left.
\frac{1}{2} [f(z)]^2 S^2 \left[\pardss W_j-\frac{\gamma}{\delta}
(\pards W_j)^2\right]+
\frac{1}{2}\beta^2 \left[\pardzz W_j-\frac{\gamma}{\delta}
(\pardz W_j)^2\right]+ \right.\nonumber  \\
\left.\beta fS\rho\left[ \pardsz W_j-\frac{\gamma}{\delta}
\pards W_j\pardz W_j \right]
\right\} =0\,. \nonumber
\eea

\section{Small transaction costs and fast mean reverting volatility:
the asymptotic analysis}

\setcounter{equation}{0}

We now suppose small transaction costs and fast mean reverting volatility.
Moreover we will assume that the transaction costs are much smaller  than
the rate of mean reversion.

$$
\lambda=\mu=\ep^2\; , \qquad \xi=\epin \; ,\qquad
\beta=\frac{\sqrt{2}\nu}{\sqrt{\ep}}\; .
$$
Buying and selling costs are assumed to be the same for simplicity.

We believe that our asymptotic assumptions are consistent with a
situation where a large investor, facing very small transaction costs,
is involved.
In fact, in the empirical study \cite{FPSS99} it is found that
$\ep\sim .005$. For large investor, typically $\lambda\lesssim .01\%$.

In absence of transaction costs  and with a deterministic
volatility $\ep=0$, the investor would  continuously trade and get
a perfect hedge staying at $y=y^*$, the ``B\&S'' hedging strategy.
When transaction costs are present there is a strip of small
thickness around $y=y^*$  where he does not transact. To resolve
this strip we introduce the inner rescaled coordinate $Y$:

\be
y=y^*+\ep^a Y\qquad \mbox{and} \qquad
\pardy\longrightarrow \ep^{-a}\pardY\,.  \label{rescaled}
\ee


\noindent The unknown boundaries between the no-transaction region
and the buy and sell regions are located at:

$$
y=y^*+\ep^a Y^+ \qquad \mbox{and} \qquad
y=y^*-\ep^a Y^-\,.
$$

\noindent It is very important from the practical hedger point of view
to determine $Y^+$ and $Y^-$.

We impose the following matching conditions (see e.g.\cite{WW}):

$$
\begin{array}{ll}
W_{NT}(Y=Y^\pm)=W(y=y^*\pm\ep^a Y^\pm)& \
\mbox{continuity}\\
&\\
\pardY W_{NT}(Y=Y^\pm)=\ep^a\pardy W(y=y^*\pm\ep^a Y^\pm)&
\mbox{ continuity of the }\\
& \mbox{ first derivative} \\
&\\
\pardYY W_{NT}(Y=Y^\pm)=
\ep^{2a}\pardyy W(y=y^*\pm\ep^a Y^\pm)&
\mbox{ smooth pasting}\\
& \mbox{ boundary condition}
\end{array}
$$

\noindent These boundary conditions will force, in the asymptotic analysis
below, $a=1/3$.
Therefore the strip of no-transaction will have a thickness
$O(\ep^{1/3})$.
\vskip.8cm

{\bf In the buy region ($ Y<Y^-$) we have the equation:}
\be
\left( \pardy W_j -(1+\lambda) S \right) =0\; ,
\ee
which solves to:
\be
W=(1+\lambda)S y+H^-(t,S,\lambda)\,.
\ee

\vskip.8cm

{\bf In the sell region ($ Y>Y^+$) we have the equation:}
$$
\left( \pardy W_j -(1-\lambda) S \right) =0 \; ,
$$
which solves to:
\be
W=(1-\lambda)S y+H^+(t,S,\lambda)\,.
\ee

{\bf In the no-transaction region we have the equation:}
\bea
\pardt W_j -r  W_j + \alpha S \pards W_j +
\epin (m-z) \pardz W_j && \nonumber \\
+\frac{1}{2} [f(z)]^2 S^2 \left[\pardss W_j-\frac{\gamma}{\delta}
(\pards W_j)^2\right]+\epin\nu^2 \left[\pardzz W_j-\frac{\gamma}{\delta}
(\pardz W_j)^2\right]&& \nonumber  \\
+\epins\nu\sqrt{2}fS\rho\left[ \pardsz W_j-\frac{\gamma}{\delta}
\pards W_j\pardz W_j \right]&=&0,
\eea

\noindent whose solution will be find in what follows.

\subsection{The solution in the no transaction  region}

As we said, in the no transaction region we use the rescaled
variable $Y$ defined by (\ref{rescaled}).
The change of variable leads to the following transformation
rules for the derivatives:
\bea
\pardy&\longrightarrow &\ep^{-1/3} \pardY \nonumber \\
\pards&\longrightarrow &\pards -\ep^{-1/3} y^*_S \pardY \nonumber \\
\pardt&\longrightarrow &\pardt -\ep^{-1/3} y^*_t \pardY \nonumber \\
\pardz&\longrightarrow &\pardz -\ep^{-1/3} y^*_z \pardY \nonumber
\eea

We write the solution in the no transaction region in the
form:

\be W_{NT}=S(y^*+\ep^{1/3}Y)+ U_0(S,t,z)+ \sum_{i=2}^{13}
\ep^{i/6}U_i(S,t,z)+ \ep^{14/6} U_{14}(S,t,z,Y)+... \label{expanW}
\ee

The derivative with respect to $t$ writes as:
\bea
\pardt W_{NT}= U_{0t}+\sum_{i=2}^{11} \ep^{i/6}U_{it}+
 \ep^{12/6}\left( U_{12t}- y^*_t U_{14Y}\right)+...
\nonumber
\eea

The derivatives with respect to $S$ writes as:
\bea
\pards W_{NT}&=& y^*+ U_{0S}+\ep^{2/6} \left(Y+U_{2S}\right) +
\sum_{i=3}^{11} \ep^{i/6}U_{iS} +
 \ep^{12/6}\left( U_{12S}- y^*_S U_{14Y}\right)+...
\nonumber \\
\pardss W_{NT}&=&U_{0SS}+
\sum_{i=2}^{9} \ep^{i/6}U_{iSS} +
 \ep^{10/6}\left( U_{10S}+ (y^*_S)^2 U_{14YY}\right)+...
\nonumber
\eea

The derivatives with respect to $z$ writes as:
\bea
\pardz W_{NT}&=& U_{0z}+
\sum_{i=2}^{11} \ep^{i/6}U_{iz} +
 \ep^{12/6}\left( U_{12z}- y^*_z U_{14Y}\right)+...
\nonumber \\
\pardzz W_{NT}&=&U_{0zz}+
\sum_{i=2}^{9} \ep^{i/6}U_{izz} + \ep^{10/6}\left( U_{10zz}+
(y^*_z)^2 U_{14YY}\right)+... \eea

The derivative with respect to $S$ and $z$ writes as:
\bea
\pardsz W_{NT}&=& U_{0Sz}+
\sum_{i=2}^{9} \ep^{i/6}U_{iSz} +
 \ep^{10/6}\left( U_{10Sz}+ y^*_Sy^*_z U_{14YY}\right)+...
 \nonumber
\eea

\vskip.8cm

\subsection{The $O(\ep^{-1})$ up to $O(\ep^{-1/6})$ order equations}

To simplify the notation, and following the use in \cite{FPSS03} and
\cite{JS}, we define the linear operators $\mathcal{L}_i$ and the
non linear operator $\mathcal{N_L}$:
\bea
\mathcal{L}_0 U&=&(m-z)U_z+\nu^2  U_{zz} \\
\mathcal{L}_1 U&=&-\nu\sqrt{2}\rho\frac{(\alpha-r)}{f} U_z
+ \nu\sqrt{2}fS\rho   U_{Sz}\\
\mathcal{L}_2 U&=&U_t+\frac{1}{2}f^2S^2U_{SS}-rU+rSU_S \\
\mathcal{N_L} U&=&-\nu^2 \frac{\gamma}{\delta}(U_{z})^2
\eea

\noindent The $O(\ep^{-1})$ equation is simply:
\be \mathcal{L}_0 U_0
+\mathcal{N_L}U_0=0 \; . \ee
The above equation can be considered
an ODE in $z$ for $U_0$:
\be \nu^2 U_{0zz}+(m-z)U_{0z}-
\nu^2\frac{\gamma}{\delta}\left(U_{0z}\right)^2=0\,. \ee

\noindent In \cite{JS}
it is proved that the only solution of an equation of this form is
a $U$ which does not depend on $z$.  The conclusion we therefore
draw is that:
$$
U_0\qquad \mbox{does not depend on $z$} \; .
$$

\noindent The $O(\ep^{-i/6})$ equations, for $i=2,\cdots ,5$ are:
$$
\mathcal{L}_0U_i=0 \; .
$$
\noindent The conclusion is:
$$
U_i\quad i=2,\cdots ,5 \qquad \mbox{does not depend on $z$} \; .
$$

\subsection{The $O(1)$ equation}

The $O(1)$ equation writes:

\bea U_{0t}  -r\left(Sy^*+U_0\right) +\alpha S
\left(y^*+U_{0S}\right)+ \mathcal{L}_0U_6 &&
\nonumber \\
+\frac{1}{2}[f(z)]^2 S^2\left[ U_{0SS}-\frac{\gamma}{\delta}\left(
y^*+U_{0S}\right)^2\right] &=&0. \label{order0temp} \eea

\noindent The above equation will be analyzed in the subsection 3.5.

\subsection{The $O(\ep^{1/6})$ equation}

The $O(\ep^{1/6})$ equation is:
$$
\mathcal{L}_0 U_7 =0 \; .
$$
We therefore get:
$$
U_7 \qquad \mbox{does not depend on $z$} \; .
$$

\subsection{The $O(\ep^{2/6})$ equation}

The $O(\ep^{2/6})$ equation is:

\bea \mathcal{L}_0 U_8 + U_{2t} -r\left(SY+U_2\right) +\alpha S
\left(Y+U_{2S}\right)+
&&\nonumber \\
+\frac{1}{2}[f(z)]^2 S^2\left[  U_{2SS}
-2\frac{\gamma}{\delta}\left(
y^*+U_{0S}\right)\left(Y+U_{2S}\right)\right]
&=&0\,.
\label{order1/3temp}
\eea

\noindent In the above equation there are terms that do not depend on $Y$,
and terms linear in $Y$. They must be equal to zero separately.
>From the terms linear in $Y$ one gets:

\be
y^*=-U_{0S} +\frac{(\alpha-r)\delta}{f^2S\gamma}
\; . \label{y*=U_0S}
\ee

\noindent The above expression gives the leading order (in absence of transaction
costs) optimal hedging strategy.
One recognizes the Black\&Scholes $\delta$-hedging strategy.

If one inserts the above expression into the $O(1)$ equation
(\ref{order0temp}), one gets:
\be
\mathcal{L}_0 U_6+
\pardt U_0+\frac{1}{2}f^2S^2\pardss U_0+rS\pards U_0
-rU_0 +\frac{1}{2}\frac{\delta}{\gamma}\frac{(\alpha-r)^2}{f^2}=0\,.
\label{order0temp2} \ee

The above equation, considered as an ODE
for $U_6$, is of the form:
\be \mathcal{L}_0 U=\chi \; .
\label{poisson} \ee
In \cite{FPSS03} it is shown that the
solvability condition for the equation (\ref{poisson}) is:

\be \left<\chi\right> =0 \; , \label{solvability} \ee

\noindent where the average $\left<\cdot\right>$ is taken with respect to
the Ornstein--Uhlenbeck process invariant measure:

\be \left<\chi\right> = \frac{1}{\nu\sqrt{2\pi}}\int_\mathbb{R}
\chi(z)e^{-(m-z)^2/2\nu^2}dz. \ee

Therefore the solvability condition for (\ref{order0temp2}) is:
\be
\pardt U_0 +\frac{1}{2}\bar{\sigma}^2S^2\pardss U_0 -rU_0+
rS\pards U_0 = -\frac{\delta(\alpha-r)^2}{2\gamma}\frac{1}{\bar{\tau}^2}
\; , \label{order0}
\ee

\noindent where $\bar{\sigma}$ is the effective constant volatility:
$$
\bar{\sigma}^2=\left<f^2\right> \; ,
$$

\noindent and $\bar{\tau}$ is defined as:
$$
\frac{1}{\bar{\tau}^2}=\left<\frac{1}{f^2}\right>\,.
$$

\noindent Once one imposes to (\ref{order0}) the appropriate final
condition, which will be different for the investor with option
liability and the investor without it, then $U_0$ is determined.
One can go back to equation (\ref{order0temp2}) and solve it for
$U_6$. The solution can be written in the form:

\be
U_6=U_6^{(z)}(S,z,t)+\tilde{U}_6(S,t) \; ,\label{express2U6}
\ee
where, $U_6^{(z)}(S,z,t)$, the part of $U_6$ which depends on
$z$, has the following expression:
$$
U^{(z)}_6(S,z,t)=-\mathcal{L}_0^{-1}\left[\frac{1}{2}S^2(f^2-\bar{\sigma}^2)U_{0SS}
+\frac{1}{2}\frac{\delta}{\gamma}(\alpha-r)^2\left(\frac{1}{f^2}-\frac{1}{\bar{\tau}^2}\right)\right]
\; ;
$$
on the other hand $\tilde{U}_6$ is a function that does not depend on $z$
and that will be determined by the $O(\ep)$ equation in the asymptotic
procedure.

We can get a more explicit representation for $U_6^{(z)}$, that
will be useful in the next subsection. We first define the
functions $\phi(z)$ and $\psi(z)$ as the solutions of the
following problems:

\bea
\mathcal{L}_0\phi&=&f^2-\left<f^2\right> \; ,\label{phi}\\
\mathcal{L}_0\psi&=&\frac{1}{f^2}-\left<\frac{1}{f^2}\right> \;
.\label{psi} \eea

\noindent Therefore the above expression for $U_6^{(z)}$ can be written as:

\be U_6^{(z)}=-\left[\frac{1}{2}S^2U_{0SS}\phi+
\frac{1}{2}\frac{\delta}{\gamma}(\alpha-r)^2\psi \right] \; .
\label{U6z} \ee

\noindent One can now go back to  equation (\ref{order1/3temp}), collect all
terms independent of $Y$ and get the following equation:
$$
\mathcal{L}_0U_8+\mathcal{L}_2 U_2=0
\label{order1/3}
$$
\noindent The above equation is a Poisson problem of the type
(\ref{poisson}). The solvability condition is:
\be
\left<\mathcal{L}_2\right> U_2=0 \; . \label{order2/6} \ee

\noindent Notice
that the above equation is homogeneous in $U_2$. Given that the
final condition, both for the investor with option liability and
for the investor without it, is $0$, one gets the following
conclusions: \bea
U_2 &\equiv& 0 \nonumber \\
U_8 &=& \tilde{U_8}(S,t) \qquad \mbox{is independent of $z$}
\nonumber \; .
\eea

\noindent Therefore the correction to the Black and Scholes value comes up to
$O(\ep^{1/2})$ order.

\subsection{The $O(\ep^{3/6})$ equation}

Using the expression (\ref{y*=U_0S}) for $y^*$, the $(\ep^{3/6})$
equation can be written as:

\be \mathcal{L}_2U_3+\mathcal{L}_1U_6+\mathcal{L}_0U_9=0\,.
\label{order1/2temp} \ee

\noindent Notice that in the above equation appears $U_6$ which, until now
we have derived only up to the function $\tilde{U}_6(S,t)$, to be
determined by to a higher order asymptotics. However, in
(\ref{order1/2temp}) $U_6$ is hit by the operator $\mathcal{L}_1$,
which cancels $\tilde{U}_6(S,t)$.

Therefore one can consider (\ref{order1/2temp}) as a Poisson
problem for $U_9$, whose solvability condition is:

\be \left<\mathcal{L}_2\right> U_3=-\left< \mathcal{L}_1U_6\right>\,.
\label{order1/2temp2} \ee

\noindent The above equation is a Black and Scholes equation for
$U_3$ with $0$ final condition and with  a source term.
We now want to rewrite the source term.

By using for $U_6$ the expression
(\ref{express2U6}) the operator $\LL_1$
cancels the part not depending on $z$, and taking into account the expression (\ref{U6z}),
one can express the source term in the equation
(\ref{order1/2temp2}) as:
\bea
-\left<\mathcal{L}_1U_6\right> &=&
\nonumber \\
\left<\mathcal{L}_1\left[\frac{1}{2}S^2U_{0SS}\phi+
\frac{1}{2}\frac{\delta}{\gamma}(\alpha-r)^2\psi \right]\right>
&=&
\nonumber \\
\frac{\nu\rho}{\sqrt{2}}\left[
\left<f\phi'\right>\left(S^3U_{0SSS}
+2S^2U_{0SS}\right)-(\alpha-r)S^2U_{0SS}\left<\frac{\phi'}{f}\right>
-\frac{\delta}{\gamma}(\alpha-r)^3\left<\frac{\psi'}{f}\right>
\right]\,.&&\nonumber \eea

\noindent Therefore $U_3$ solves the following Black and Scholes equation:
\bea
U_{3t} +\frac{1}{2}\bar{\sigma}^2S^2U_{3SS}-rU_3 +rSU_{3S}&=&
\nonumber \\
\frac{\nu\rho}{\sqrt{2}}\left[
\left<f\phi'\right>\left(S^3U_{0SSS}
+2S^2U_{0SS}\right)-(\alpha-r)S^2U_{0SS}\left<\frac{\phi'}{f}\right>
-\frac{\delta}{\gamma}(\alpha-r)^3\left<\frac{\psi'}{f}\right>
\right]&&\nonumber \\
\label{order1/2}
\eea
with zero final data.

\subsection{The $O(\ep^{2/3})$ equation}

Since $U_7$ does not depend on $z$, the
$O(\ep^{2/3})$ equation can be written as:

\be \mathcal{L}_2U_4 +\mathcal{L}_0U_{10}+\nu^2(y^*_z)^2U_{14YY}
-\frac{\gamma}{\delta}f^2S^2Y^2 =0 \; . \label{order4/6temp} \ee
The equation (\ref{order4/6temp}) can be considered as an ODE in
$Y$ for $U_{14}$. It writes as:

\be U_{14YY}=A Y^2 + B \; ,
\label{ODE} \ee

\noindent where we have defined the following quantities:

\begin{eqnarray*}
A&=&\frac{\gamma}{\delta}\frac{f^2 S^2 } {\nu^2
\left(y^*_z\right)^2}\,,\\
B&=&-\frac{\mathcal{L}_2U_4+\mathcal{L}_0U_{10}}
{\nu^2 \left(y^*_z\right)^2}  \; .
\end{eqnarray*}

\noindent The above equation \eqref{ODE} solves to:
\be
U_{14} = \frac{A}{12}Y^4 + \frac{1}{2} B Y^2 + C Y +D \,,\label{U14}
\ee
with $C$ and $D$ independent of $Y$. Now we have to impose the matching conditions.

Being:
\be
\begin{array}{lll}
W_{BUY}=&(1+\ep^2)S y  +H^-(S,z,t)&\quad\mbox{in the outer}\\
&&\quad\mbox{buy region}
 \\ \\
W_{SELL}=&(1-\ep^2)S y  +H^+(S,z,t)&\quad\mbox{in the outer}\\
&&\quad\mbox{sell region}
\end{array}
\ee
\noindent and imposing the continuity of the gradient at the two boundaries:
\bea
\pardY W_{NT}(Y=-Y^-)= \ep^{1/3}\pardy W_{BUY}(y=y^*-\ep^{1/3}Y^-)\,,
\nonumber \\
\pardY W_{NT}(Y=Y^+)= \ep^{1/3}\pardy W_{SELL}(y=y^*+\ep^{1/3}Y^+)\,,
\nonumber \eea

\noindent one gets, at the $O(\ep^{14/6})$: \bea
\pardY U_{14} (Y=-Y^-) = S\,, \\
\pardY U_{14} (Y=Y^+) =-S\,.
\eea
\noindent
Therefore, using (\ref{U14}), one gets:
\bea
-\frac{A}{3} \left(Y^-\right)^3  -B Y^- +C &=&S\,,  \label{cubic1}  \\
\frac{A}{3} \left(Y^+\right)^3 + B Y^+ +C  &=&-S \,.   \label{cubic2}
\eea

\noindent Moreover, being $W$ in the outer regions linear in $y$, one imposes the continuity of the second derivative as follows:
$$
\pardYY W_{NT} (Y=\pm Y^\pm)=0\,,
$$
\noindent i.e.:
\bea
A(Y^+)^2+B =0\,, \nonumber \\
A(Y^-)^2+B=0\,.  \nonumber \eea

\noindent From these equations one sees that,
at this order, the bandwidth about the Black and Scholes strategy
is symmetric, i.e.: \be Y^+=Y^-=\left(-\frac{B}{A}\right)^{1/2}
\qquad \ee Subtracting the two equations (\ref{cubic1}) and
(\ref{cubic2}) to eliminate $C$, and using the above expressions
for $Y^\pm$, one gets:
$$
\frac{2}{3}(-B)^{3/2} A^{-1/2} =S\,.
$$
After some manipulations, and using the expressions for $A$ and
$B$, the equation (\ref{ODE}) leads to the following equation: \be
\mathcal{L}_0U_{10} +\mathcal{L}_2U_4 =
\left[\frac{3}{2}fS^2\sqrt{\frac{\gamma}{\delta}}
\nu^2(y^*_z)^2\right]^{2/3} \label{oreder2/3temp2} \ee

One can also find an expression for the amplitude of the
no-transaction region:
\be
Y^+=Y^-=
\left[ \frac{3}{2} \frac{1}{f^2S}\frac{\delta}{\gamma}
\nu^2(y^*_z)^2\right]^{1/3}  \;  .  \label{boundaries}
\ee

Equation (\ref{oreder2/3temp2}) is a Poisson problem of the type
of equation (\ref{poisson}). The solvability condition gives an
equation for $U_4$:
\be \left< \mathcal{L}_2\right> U_4=
\left<\left[\frac{3}{2}fS^2\sqrt{\frac{\gamma}{\delta}}
\nu^2(y^*_z)^2\right]^{2/3}\right> \,.\label{order2/3} \ee

\noindent Notice also that, adding the two equations (\ref{cubic1}) and
(\ref{cubic2}), one gets that $C=0$. Therefore:
\be U_{14} =
\frac{A}{12}Y^4 + \frac{1}{2} B Y^2  +D \,,\label{U14final} \ee

\noindent which will be useful in the subsection 3.9.

\subsection{The $O(\ep^{5/6})$ equation}

The $O(\ep^{5/6})$ equation writes as:

\be \mathcal{L}_0 U_{11}+\mathcal{L}_2 U_5-\sqrt{2}\frac{\partial
U_6}{\partial z} \frac{\gamma}{\delta}\nu\rho f(z)
SY-\frac{\partial U_3}{\partial
S}\frac{\gamma}{\delta}f(z)^2S^2Y+\frac{\partial^2
U_{15}}{\partial Y^2}\left(\frac{\partial y^*}{\partial
z}\right)^2\nu^2=0\,, \label{order5}\ee

This equation can be considered as an ODE for $U_{15}$:

$$\frac{\partial^2
U_{15}}{\partial Y^2}=\bar{A}Y+\bar{B}\,,$$

\noindent where we have denoted:

\begin{eqnarray*}\bar{A}&=&\left(\sqrt{2}\frac{\partial U_6}{\partial z}
\frac{\gamma}{\delta}\nu\rho f(z) S-\frac{\partial U_3}{\partial
S}\frac{\gamma}{\delta}f(z)^2S^2\right)/(\nu^2y^{*2}_z)\,,\\
\bar{B}&=&-\frac{\mathcal{L}_0 U_{11}+\mathcal{L}_2
U_5}{\nu^2y^{*2}_z}\,.
\end{eqnarray*}

\noindent Integrating (\ref{order5}) with respect to $Y$ and using the
boundary conditions:

$$ U_{15Y}(Y^+)=U_{15Y}(-Y^-)=0 \,$$

\noindent which are needed to ensure the continuity of the gradient, one
gets:

\be\mathcal{L}_0 U_{11}+\mathcal{L}_2 U_5=0 \,.\ee

\noindent The above equation is a Poisson problem for
$U_{11}$, whose solvability condition reads:

\be <\mathcal{L}_2> U_5=0\,. \label{u5eq}\ee

\noindent This is a homogeneous Black-Scholes equation for $U_5$. Given that
the final condition is zero, we gets the conclusions:

\bea
U_5 &\equiv& 0\,, \nonumber \\
U_{11} &=& \tilde{U_{11}}(S,t) \qquad \mbox{does not depend on
$z$} \nonumber \; . \eea

\subsection{The $O(\ep)$ equation}

Collecting the $O(\ep)$ terms one gets:
\bea\label{order1temp}
\mathcal{L}_2U_6+\mathcal{L}_1U_9+\mathcal{L}_0U_{12}
-\frac{1}{2}f^2S^2\frac{\gamma}{\delta}(U_{3S})^2
-\nu\sqrt{2}fS\rho\frac{\gamma}{\delta}U_{3S}U_{6z}-y^*_z(m-z)U_{14Y}
&&\\-\frac{\gamma}{\delta}f^2YS^2U_{4S} +\nu^2\left[
-y^*_{zz}U_{14Y}-2y^*_zU_{14Yz}+(y^*_z)^2U_{16YY}
-\frac{\gamma}{\delta}(U_{6z})^2\right] \nonumber&=&0\,.
 \eea

\noindent The above equation can be considered as an ODE in $Y$ for
$U_{16}$:

$$\frac{\partial^2 U_{16}}{\partial Y^2}=\tilde{A}Y+\tilde{B},$$

\noindent where we have defined:

\begin{eqnarray*}
\tilde{A}&=&\frac{\gamma}{\delta}\frac{f^2S^2U_{4S}}{ \nu^2
(y^*_z)^2}\,,\\
\tilde{B}&=&-\left[\mathcal{L}_0 U_{12}+\mathcal{L}_1
U_9+\mathcal{L}_2 U_6-\frac{1}{2}f^2S^2
\frac{\gamma}{\delta}U_{3S}^2-\nu \sqrt{2}f S\rho
\frac{\gamma}{\delta}
U_{3S}U_{6z}+\right.\\&-&\left.\nu^2\left(y^*_{zz}U_{14Y}+2y_z^*U_{14Yz}+\frac{\gamma}{\delta}
U_{6z}^2\right)+U_{14Y}y^*_z(m-z)\right]/\left(\nu^2
(y_z^*)^2\right)\,.
\end{eqnarray*}

\noindent Notice that $\tilde{A}$ does not depend on $Y$ and in $\tilde{B}$
the $Y$-dependent terms appear only with their derivatives in $Y$.

We integrate the equation (\ref{order1temp}) from $-Y^-$ and
$Y^+$.

Let us use the boundary conditions:
$$
U_{16Y}(Y^+)=U_{16Y}(-Y^-)=0 \;.
$$

\noindent  Moreover, being $Y^+=Y^-$ and from the expression (\ref{U14final}) it follows that:
$$
\int^{Y^+}_{-Y^-}U_{14Y}dY=0\;.
$$

\noindent Integrating (\ref{order1temp}) we get:

\be
\mathcal{L}_0U_{12}+\mathcal{L}_1U_9+\mathcal{L}_2U_6
=\frac{1}{2}f^2S^2\frac{\gamma}{\delta}(U_{3S})^2
+\nu\sqrt{2}fS\rho\frac{\gamma}{\delta}U_{3S}U_{6z}
+\nu^2\frac{\gamma}{\delta}(U_{6z})^2\,.
\nonumber
\ee

\noindent The solvability condition for $U_{12}$, gives the following equation for
$U_6$:
\be \left<\mathcal{L}_2\right>U_6=
-\left<\mathcal{L}_1U_9\right>
+\frac{1}{2}\bar{\sigma}^2S^2\frac{\gamma}{\delta}(U_{3S})^2
+\nu\sqrt{2}S\rho\frac{\gamma}{\delta}U_{3S}\left<fU_{6z}^{(z)}\right>
+\nu^2\frac{\gamma}{\delta}\left<(U_{6z}^{(z)})^2\right>\,.
\label{order1} \ee

\noindent The main results of this section are the following:
\begin{enumerate}
\item\noindent Equation (\ref{order0}) for $U_0$;
\item\noindent Equation (\ref{order2/6}) for $U_2$ which led us to $U_2\equiv0$
\item\noindent Equation (\ref{order1/2})  for $U_3$;
\item\noindent Equation (\ref{order2/3}) for $U_4$;
\item\noindent Equation (\ref{u5eq}) for $U_5$ which led us to $U_5\equiv0$
\item\noindent Equation (\ref{order1}) for $U_6$;
\item\noindent The expression (\ref{y*=U_0S}) for $y^*$, the
center of the no-transaction region.
\item\noindent The expression (\ref{boundaries}) for the boundaries of the no
transaction region.
\end{enumerate}

\section{The price of the option}
\setcounter{equation}{0}

To calculate the price of the option we now use the equation
(\ref{price2}). The price will have the same asymptotic
expansion as $W_j$ with $j=1,w$, namely:

\be C=C_0+\ep^{1/3}C_2+\sqrt{\ep}C_3+\ep^{2/3}C_4+\ep^{5/6}C_5+\ep
C_6 \; . \label{priceasympt} \ee Each $C_i$ is given by:
$$
C_i=U^1_i-U^w_i \; .
$$
To find the appropriate final conditions for the $C_i$, we write
the final conditions for $W^1$ and $W^w$. They are, respectively:
\be W^1(T)=y(T)S(T) \ee and \be W^w(T)=y(T)S(T)-\max(S(T)-K,0) \ee
Given the expression (\ref{expanW}) one has that the final
conditions for the the $U_i$ are the following:

\be U^1_i(T)=0 \qquad\mbox{for}\quad i=0,2,3,4,5,6.
\label{finconU^1} \ee \bea
U^w_0(T)&=&-\max(S(T)-K,0) \label{finconU^w_0}\\
U^w_i(T)&=&0 \qquad\mbox{for}\quad i=2,3,4,5,6.
\label{finconU^w_i}\nonumber \eea

\subsection{The leading order price}

To calculate the leading order price we have to calculate
$U^1_0$ and $U^w_0$ where they both satisfy equation (\ref{order0}).
Given the respective final conditions (\ref{finconU^1}) and
(\ref{finconU^w_0}) one has that:
\bea
U^1_0&=&(T-t)\frac{\delta(\alpha-r)^2}{2\gamma}\frac{1}{\bar{\tau}^2}\,,
\label{U01} \\
U^w_0&=&(T-t)\frac{\delta(\alpha-r)^2}{2\gamma}\frac{1}{\bar{\tau}^2}-
C^{BS}\,, \label{U0w} \eea

\noindent where $C^{BS}$ is the classical pricing
formula for a European call option, i.e.:
$$
C^{BS}(S,t)=SN(d_1)-Ke^{-r(T-t)}N(d_2)\; ,
$$
\noindent where:
$$
d_1=\frac{\log(S/K)+(r+\frac{1}{2}\bar{\sigma}^2)(T-t)}
{\bar{\sigma}\sqrt{T-t}} \qquad d_2=d_1-\bar{\sigma}\sqrt{T-t}\; ,
$$
\noindent and $N(z)$ is the normal cumulative distribution function.
>From the above expressions for $U^j_0$ one obtains:
$$C_0(S,t)=C^{BS}(S,t) \; .$$

\subsection{The $O(\ep^{1/3})$ correction}

The equation for $U^1_2$ and $U^w_2$ is (\ref{order2/6}), a homogeneous
Black and Scholes equation.
In both cases the final condition is homogeneous.
Therefore $U^1_2\equiv 0$ and $U^w_2\equiv 0$ and:
$$
C_2(S,t)=0   .
$$

\subsection{The $O(\ep^{1/2})$ correction}

The equation for $U^1_3$ and $U^w_3$ is equation (\ref{order1/2}),
in both cases with homogeneous final condition. Using respectively
the expressions (\ref{U01}) and (\ref{U0w}) in (\ref{order1/2}),
one has that:

\begin{eqnarray}\label{U31}
U^1_3&=&(T-t)\frac{\nu
\rho}{\sqrt{2}}\frac{\delta}{\gamma}(\alpha-r)^3\left<\frac{\psi'}{f}\right>,\\\nonumber
U^w_3&=&-(T-t)\frac{\nu
\rho}{\sqrt{2}}\left[-\frac{\delta}{\gamma}(\alpha-r)^3\left<\frac{\psi'}{f}\right>-\left<f\varphi'\right>
\left(S^3 C^{BS}_{3S}+2S^2C^{BS}_{SS}\right)+\right.\\\label{U3w}
&\
&\left.+(\alpha-r)S^2\left<\frac{\varphi'}{f}\right>C^{BS}_{SS}\right].
\end{eqnarray}

\noindent Therefore:

\bea C_3(S,t)=-(T-t) \frac{\nu\rho}{\sqrt{2}}\left[
\left<f\phi'\right>\left(S^3\pardsss C^{BS} +2S^2\pardss C^{BS}
\right) -(\alpha-r)S^2\pardss
C^{BS}\left<\frac{\phi'}{f}\right>\right]&& \label{C3} \eea

\noindent The derivatives with respect to $S$ of $C^{BS}$ are explicitly
calculated in the Appendix C. Moreover, the values of
$\left<f\phi'\right>$ and $\left<\phi'/f\right>$ are computed in
Appendix B by using the Scott's model.

\subsection{The $O(\ep^{2/3})$ correction}

The equation for $U_4^1$ and $U_4^w$ is equation (\ref{order2/3}),
in both cases with homogeneous final condition.
The source term for the two problems is the same: in fact
$y_z^*$ has the same expression for both problems.
Therefore $U_4^w=U_4^1$ and
$$
C_4(S,t)=0 \; .
$$

\subsection{The $O(\ep^{5/6})$ correction}

The equation for $U_5^1$ and $U_5^w$ is the homogeneous Black and
Scholes equation (\ref{u5eq}) with zero final condition in both
cases. Then:

$$C_5(S,t)=0.$$

\noindent The correction to the price without transaction
costs with stochastic volatility comes up to $O(\ep)$.

\subsection{The $O(\ep)$ correction}

To compute the $O(\ep)$ correction we have to solve the equation
(\ref{order1}). It is a Black and Scholes equation with a source
term. We know that $U_6$ is decomposed in a part dependent on $z$
and a part that does non depend on $z$, see equation
(\ref{express2U6}). The same decomposition holds also for $C_6$:
$$
C_6=C_6^{(z)}+\tilde{C_6} \; ,
$$

\noindent where:
$$
C_6^{(z)}=U_6^{(z)1}-U_6^{(z)w}\; \; ,
$$
\noindent and
$$
\tilde{C}_6=\tilde{U}_6^{(z)1}-\tilde{U}_6^{w}\; \; .
$$

\noindent We have already computed $U_6^{(z)j}$, as given in equation
(\ref{U6z}), we can therefore calculate $C_6^{(z)}$. In fact,
using (\ref{U6z}) and the expressions (\ref{U01}) and
(\ref{U0w}), one gets:
\bea
U_6^{(z)1}&=&-\frac{1}{2}\frac{\delta}{\gamma}(\alpha-r)^2\psi\\
U_6^{(z)w}&=&\frac{1}{2}S^2C^{BS}_{SS}\varphi-\frac{1}{2}\frac{\delta}{\gamma}(\alpha-r)^2\psi
\eea

\noindent which gives:
 \be C_6^{(z)}=-\frac{1}{2}S^2C^{BS}_{SS}\phi \; . \ee

\noindent We are now left with the task of computing $\tilde{C}_6$.

The equation for $C_6$ can be derived using equation
(\ref{order1}). Subtracting the two equations relative to
$U_6^{1}$ and $U_6^{w}$ one gets:

\bea \label{Ptilde6}
\left<\mathcal{L}_2\right>C_6&=&-\left[\left<\mathcal{L}_1U_9^1\right>-\left<\mathcal{L}_1U_9^w\right>\right]
+\frac{1}{2}S^2\bar{\sigma}^2\frac{\gamma}{\delta}\left[\left(U^1_{3S}\right)^2-\left(U^w_{3S}\right)^2\right]+\\\nonumber
&+&\nu
\sqrt{2}S\rho\frac{\gamma}{\delta}\left[U^1_{3S}\left<fU^{(z)1}_{6z}\right>-U^w_{3S}\left<fU^{(z)w}_{6z}\right>\right]
+\nu^2\frac{\gamma}{\delta}\left[\left<\left(U_{6z}^{(z)1}\right)^2\right>-\left<\left(U_{6z}^{(z)w}\right)^2\right>\right] \\
&& \nonumber \eea

\noindent This equation is a Black and Scholes equation for $C_6$ with
source term. In the Appendix A this source term is explicitly
computed and the equation (\ref{Ptilde6}) writes as:

\begin{eqnarray}
\left<\mathcal{L}_2\right>C_6=(T-t)^2
\hat{A}+(T-t)\hat{B}+\hat{C}\,,\label{u6eq}
\end{eqnarray}

\noindent where:

\begin{eqnarray*}
\hat{A}&=&-\frac{\nu^2\rho^2}{4}\frac{\gamma}{\delta}S^2\bar{\sigma}^2\left[\left<f\varphi'\right>\left(S^3C^{BS}_{4S}+
5S^2C^{BS}_{SSS}+4SC^{BS}_{SS}\right)+\right.\\
&\ &
\left.-(\alpha-r)\left<\frac{\varphi'}{f}\right>\left(S^2C^{BS}_{SSS}+2SC^{BS}_{SS}\right)\right]^2\,,\\
\hat{B}&=&-\nu^2\rho^2S^2\left(\left<\varphi'f\right>-\frac{1}{2}(\alpha-r)\left<\frac{\varphi'}{f}\right>\right)
\left[\left<f\varphi'\right>\left(S^3C^{BS}_{5S}+8S^2C^{BS}_{4S}\right.\right.\\
&+&\left.\left.14SC^{BS}_{SSS}
+4C^{BS}_{SS}\right)-(\alpha-r)\left<\frac{\varphi'}{f}\right>\left(S^2C^{BS}_{4S}
+4SC^{BS}_{SSS}+2C^{BS}_{SS}\right)\right]\\
&\
&-\frac{\nu^2\rho^2}{2}S^3\left<f\varphi'\right>\left[\left(S^3C^{BS}_{6S}+11S^2C^{BS}_{5S}+30SC^{BS}_{4S}+
18C^{BS}_{SSS}\right)\left<f\varphi'\right>\right.\\
&\
&\left.-(\alpha-r)\left(S^2C^{BS}_{5S}+6SC^{BS}_{4S}+6C^{BS}_{SSS}\right)
\left<\frac{\varphi'}{f}\right>\right]\\
&\ &-
\frac{\nu^2\rho^2}{2}\frac{\gamma}{\delta}S\left[\left<f\varphi'\right>\left(S^3C^{BS}_{4S}+
5S^2C^{BS}_{3S}+4SC^{BS}_{SS}\right)\right.+\\
&-&\left.(\alpha-r)\left<\frac{\varphi'}{f}\right>\left(2SC^{BS}_{SS}+S^2C^{BS}_{3S}\right)\right]\times\\
&\times&\left(S^2C^{BS}_{SS}\left<\varphi'f\right>-\frac{\delta}{\gamma}(\alpha-r)^2\left<\psi'f\right>\right)\,,\\
\hat{C}&=&\nu^2\frac{\gamma}{\delta}\left[-\frac{1}{4}S^4\left(C_{SS}^{BS}\right)^2\left<\varphi'^{2}\right>+
\frac{1}{2}\frac{\delta}{\gamma}(\alpha-r)^2S^2C_{SS}^{BS}\left<\varphi'\psi'\right>\right]\\
&\ & -\nu^2\rho^2(\alpha-r)\left[(\alpha-r)S^2C^{BS}_{SS}
\left<\frac{G'}{f}\right>-\left(S^3C^{BS}_{SSS}+2S^2C^{BS}_{SS}\right)\left<\frac{F'}{f}\right>\right]\\
& \ &
-\nu^2\rho^2\left[\left(S^4C^{BS}_{4S}+5S^3C^{BS}_{SSS}+4S^2C^{BS}_{SS}\right)
\left<F'f\right>\right.\\
&\
&\left.-(\alpha-r)\left(S^3C^{BS}_{SSS}+2S^2C^{BS}_{SS}\right)\left<G'f\right>\right]\,.
\end{eqnarray*}

\noindent Given the homogeneous final condition, the solution of
(\ref{u6eq}) writes as:

\begin{eqnarray*}
C_6=\frac{(T-t)^3}{3}\hat{A}+\frac{(T-t)^2}{2}\hat{B}+(T-t)\hat{C}.
\end{eqnarray*}

\noindent We have used the fact that:

\begin{eqnarray*}
\mathcal{L}_2\left(\frac{(T-t)^3}{3}A+\frac{(T-t)^2}{2}B+(T-t)C\right)&=&
(T-t)^2A+(T-t)B+C+\\
&\
&+\frac{(T-t)^3}{3}\mathcal{L}_2A+\frac{(T-t)^2}{2}\mathcal{L}_2B+(T-t)\mathcal{L}_2C
\end{eqnarray*}

\noindent and the last three terms are zero:

\begin{eqnarray*}
\mathcal{L}_2\left(S^n\frac{\partial^n C^{BS}}{\partial
S^n}\right)=S^n \frac{\partial^n }{\partial
S^n}\mathcal{L}_2C^{BS}=0\,.
\end{eqnarray*}

In Appendix C the reader can find the derivatives of $C^{BS}$ with
respect to $S$ up to the sixth order. In Appendix B the
averages $<\cdot>$ with respect to the Ornstein-Uhlenbeck process
invariant measure are explicitly computed using the Scott's model.

\section{Numerical results}

In this section we present the main results obtained via the
asymptotic method. At first we plot the no transaction region for
different values of the volatility, ranging from $5\%$ up to
$60\%$, both in the case which does not include the option, denoted
by the index 1, and in the case which includes the option, denoted
by the index $w$. The volatility is chosen as in the Scott model,
$f(z)=e^z$. In figure 1 the curves representing the Black and
Scholes strategy $y^*$ in the absence of transaction costs and the
hedging boundaries, $y=y^*\pm \varepsilon^{\frac{1}{3}} Y^+$, are
plotted versus $S$ for the first problem. From the expressions
(\ref{y*=U_0S}) and (\ref{boundaries}) it follows that these
curves are respectively given by:

\bea y^*&=& \frac{(\alpha-r)\delta}{e^{2z}S\gamma}\,,\label{stra1}\\
y&=&\frac{(\alpha-r)\delta}{e^{2z}S\gamma}\pm\varepsilon^{\frac{1}{3}}\left[\frac{6(\alpha-r)^2\nu^2\delta^3}{e^{6z}S^3\gamma^3}
\right]^{\frac{1}{3}}\,.\label{front1} \eea

\noindent The corresponding curves in the second case are plotted in figure
2 and their equations are:

\bea y^*&=&C^{BS}_S+\frac{(\alpha-r)\delta}{e^{2z}S\gamma}\,,\label{stra2}\\
y&=&C^{BS}_S+\frac{(\alpha-r)\delta}{e^{2z}S\gamma}\pm\varepsilon^{\frac{1}{3}}\left[\frac{6(\alpha-r)^2\nu^2\delta^3}{e^{6z}S^3\gamma^3}
\right]^{\frac{1}{3}}\,.\label{front2} \eea

\noindent Both in the figures 1 and 2 the strike price is $K=0.5$, the
risk-free interest rate is $r=0.07$, the drift rate of the stock
is $\alpha=0.1$, the risk aversion is $\gamma=1$, the mean
volatility $\bar{\sigma}=0.2$, the time to expiry is $0.3$ and
$\varepsilon=1/200$.

\begin{figure}[h]
\begin{center}
\subfigure[] {\epsfxsize=2.5 in \epsfbox{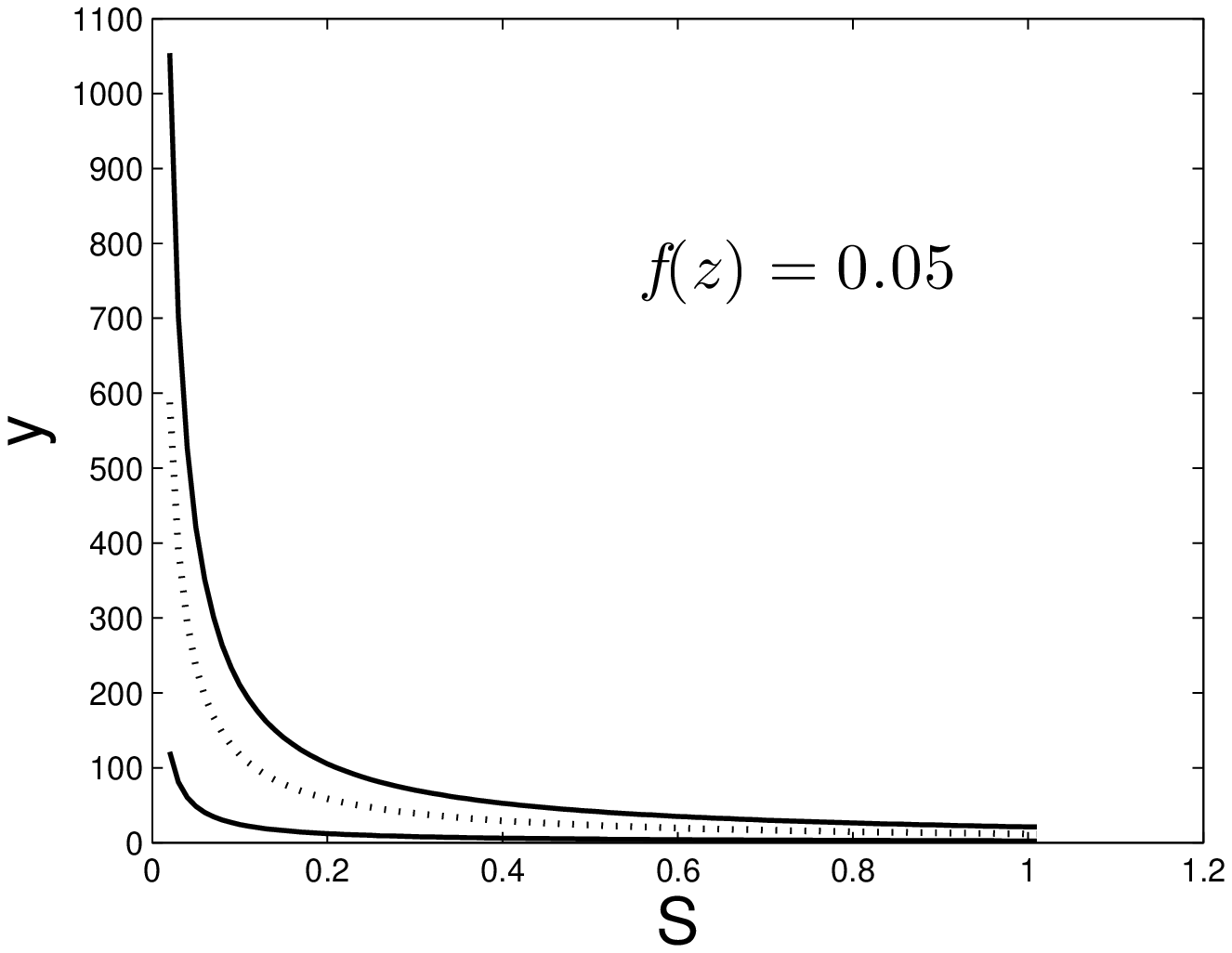}}
\subfigure[] {\epsfxsize=2.5 in \epsfbox{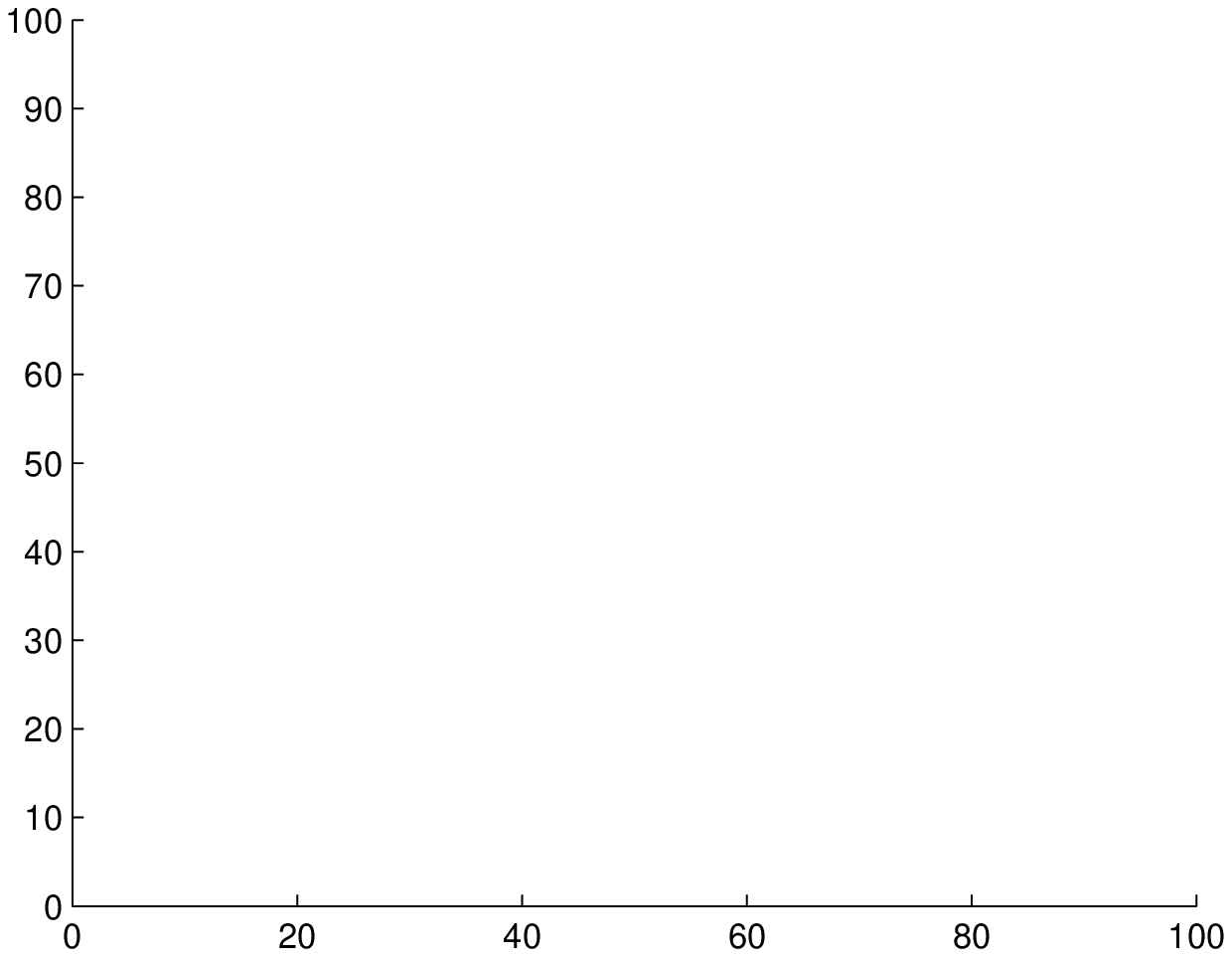}}
\subfigure[] {\epsfxsize=2.5 in \epsfbox{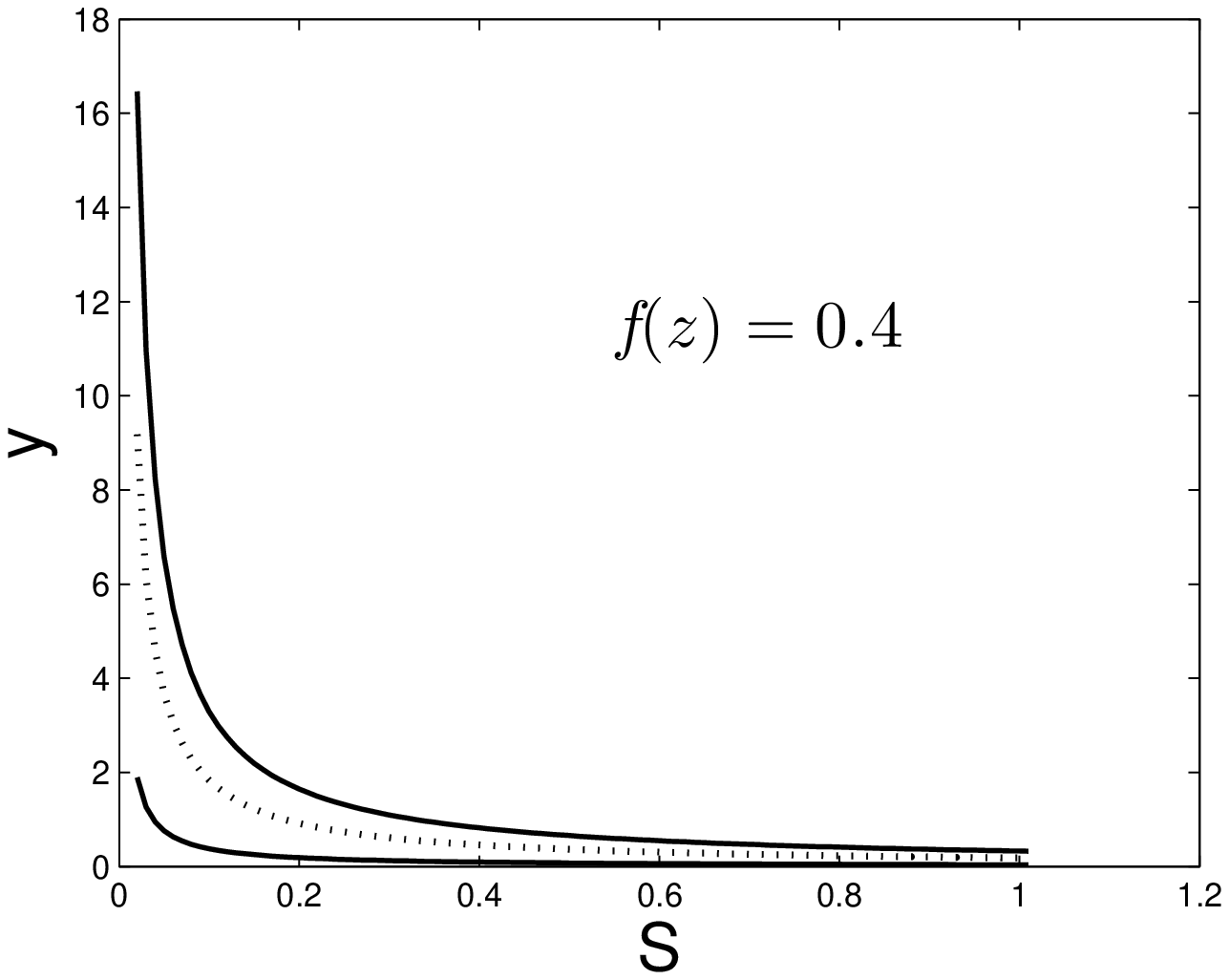}}
\subfigure[] {\epsfxsize=2.5 in \epsfbox{vuoto.eps}}
\end{center}
\caption{\label{f1} The no-transaction region in the case which
does not include the option. The dotted curve represents the Black
and Scholes strategy $y^*$ in the absence of transaction costs,
the other two curves represent the hedging boundaries. See the
text for the choice of parameters.}
\end{figure}

\begin{figure}[h]
\begin{center}
\subfigure[] {\epsfxsize=2.5 in \epsfbox{vuoto.eps}}
\subfigure[] {\epsfxsize=2.5 in \epsfbox{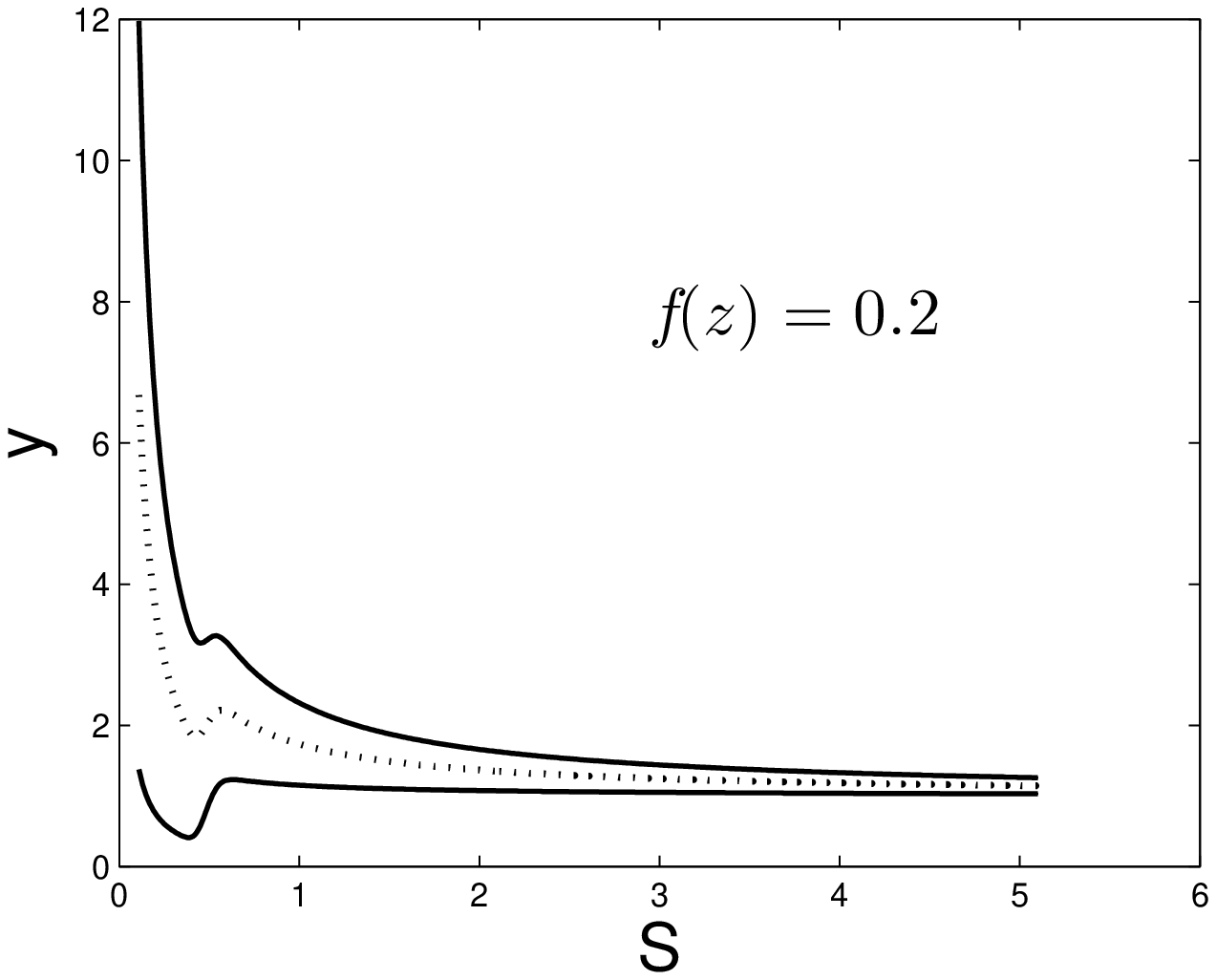}}
\subfigure[] {\epsfxsize=2.5 in \epsfbox{vuoto.eps}}
\subfigure[] {\epsfxsize=2.5 in \epsfbox{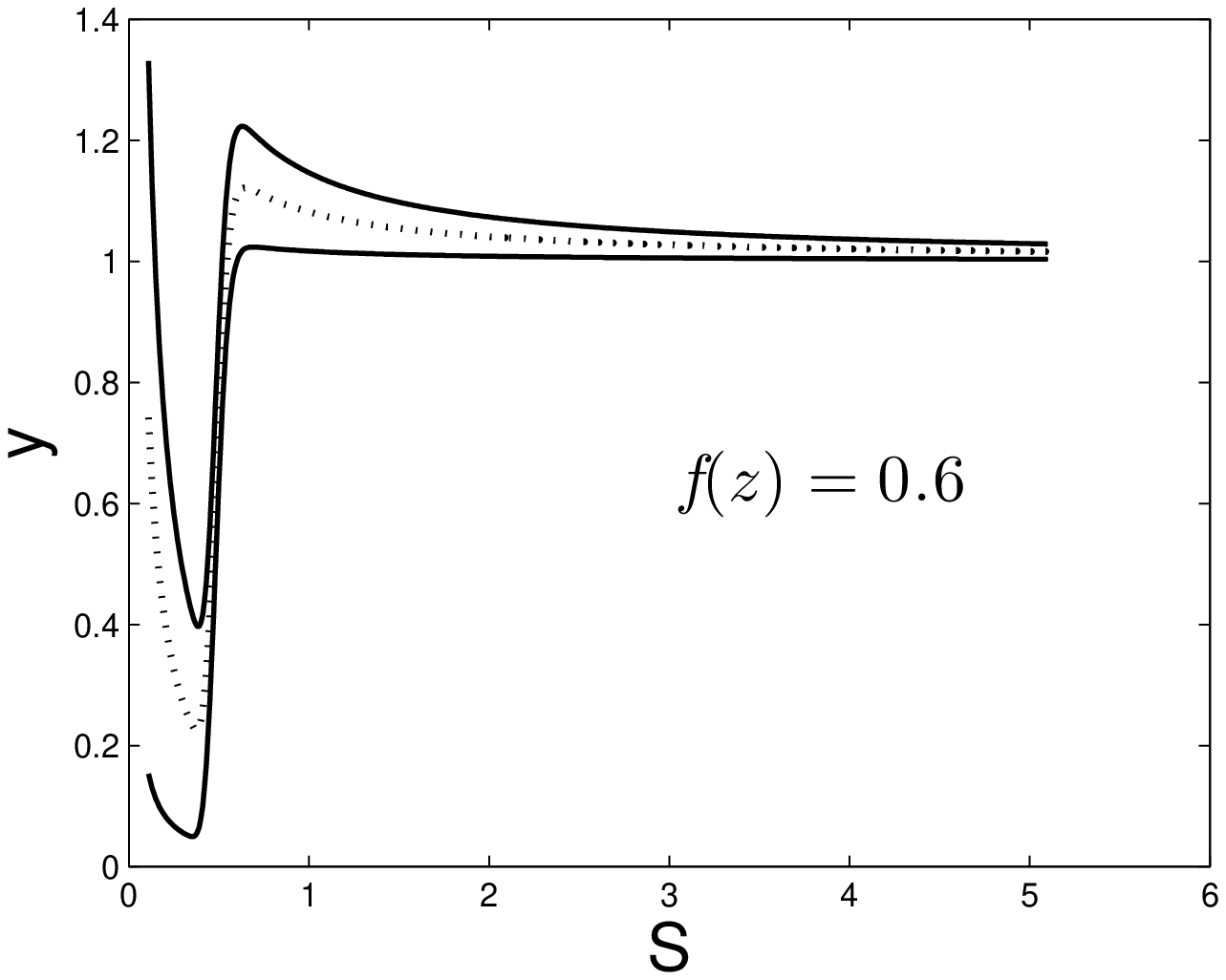}}
\end{center}
\caption{\label{f2} The no-transaction region in the case which
includes the option. The dotted curve represents the Black and
Scholes strategy $y^*$ in the absence of transaction costs, the
other two curves represent the hedging boundaries. See the text
for the choice of parameters.}
\end{figure}

Finally, in figure 3 it is shown the curve representing the
classical Black and Scholes price of a European call option with
the first correction obtained at $O(\varepsilon^{\frac{1}{2}})$
and the second correction obtained at  $O(\varepsilon)$. Here the
parameters are chosen as $K=100$, $r=0.04$, $\alpha=0.1$,
$\gamma=1$, $\bar{\sigma}=0.165$, the time to expiry is $3$,
$\varepsilon=1/200$. In figure 3(a) the correlation coefficient is
$\rho=0$, in figure 3(b) is $\rho=-0.2$.

\begin{figure}[h]
\begin{center}
\subfigure[] {\epsfxsize=2.5 in \epsfbox{vuoto.eps}} \subfigure[]
{\epsfxsize=2.5 in \epsfbox{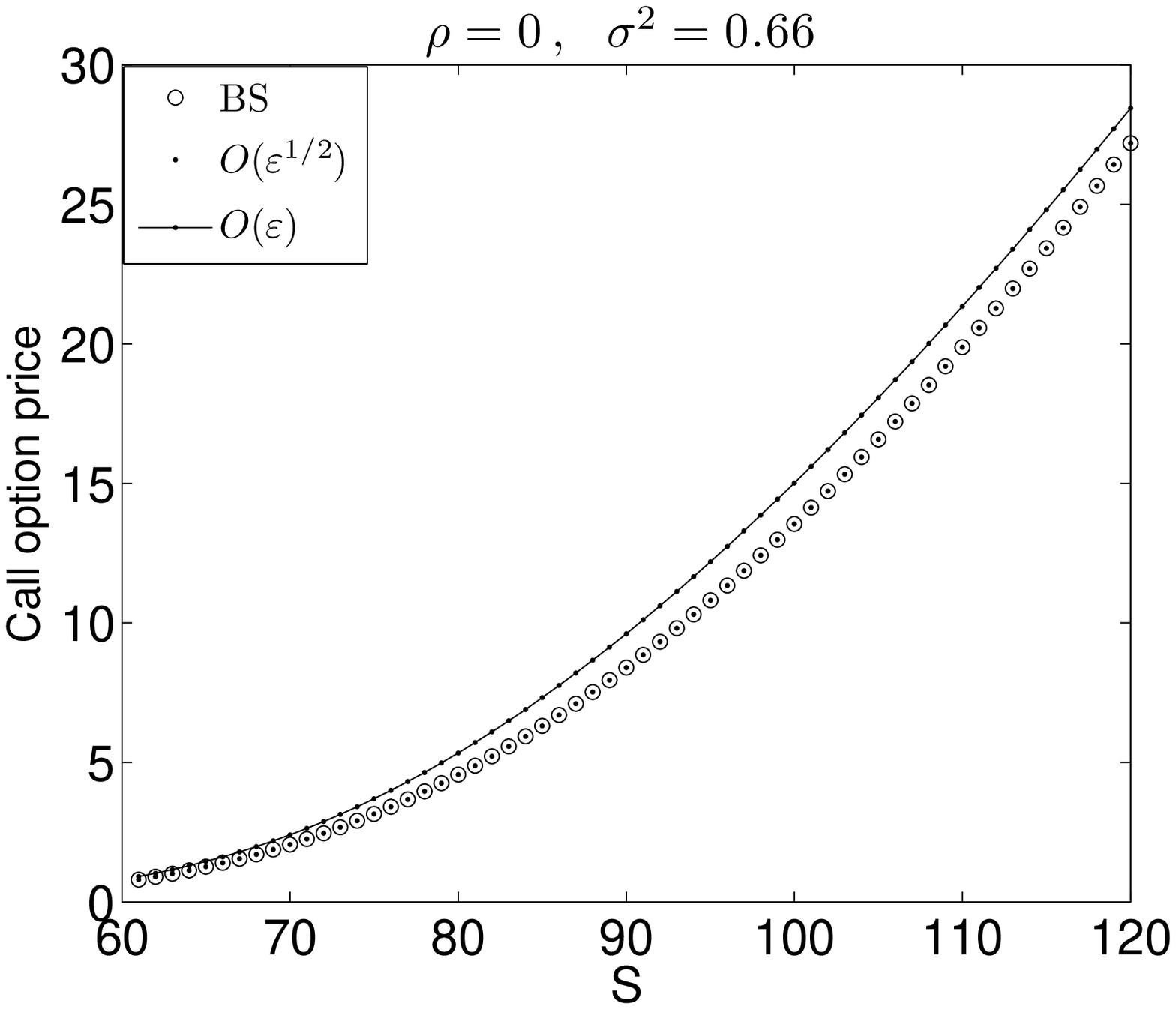}}
\end{center}
\caption{\label{f3} The dotted line represents the classical price
for a European call option; the dashed line represents the price
including the first correction at $O(\varepsilon^{\frac{1}{2}})$;
the continuous line represents the price including the first
correction and the second correction $O(\varepsilon)$. (a)
$\rho=0$; (b) $\rho=-0.2$.}
\end{figure}

We want to remark that the oscillatory behavior exhibited in figure
2b), 2c), 2d) by $y$ for small values of $S$ has been already observed by A.E.Whalley
and P.Wilmott \cite{WW} also in models with constant volatility,
although this feature seems to be more pronounced in the present context.
Moreover the thickness of the no transaction region in presence of stochastic
volatility seems to be bigger than in model with constant volatility.
The $(\epsilon)^{1/3}$ scaling was also a relevant feature already established in \cite{WW}
which is exhibited also by the present model.

\appendix

\section{The source term in the equation for $C_6$}

\end{document}